\documentclass[preprint,12pt]{elsarticle}

\usepackage{booktabs} 
\usepackage{algorithm}
\usepackage{algorithmic}
\usepackage{amsmath}
\usepackage{graphicx}
\usepackage{epstopdf}
\usepackage{amsfonts}

\journal{Neurocomputing}

\begin{document}

\begin{frontmatter}



\title{NEW: A Generic Learning Model for Tie Strength Prediction in Networks}


\author{Zhen Liu\corref{Zhen Liu}\fnref{1,2}}
\ead{quake.liu0625@gmail.com}
\ead{quake@uestc.edu.cn}
\cortext[Zhen Liu]{Corresponding author is Zhen Liu and both Zhen Liu and Hu Li are co-first authors for their equal contributions to this work.}
\author{Hu Li\fnref{1}}
\author{Chao Wang\fnref{3}}

\address[1]{Web Sciences Center, School of Computer Science and Engineering, University of Electronic Science and Technology of China, Chengdu, China}
\address[2]{Big Data Research Center, School of Computer Science and Engineering, University of Electronic Science and Technology of China, Chengdu, China}
\address[3]{School of Electrical Engineering, ChongQing University, Chongqing, China}

\begin{abstract}
Tie strength prediction, sometimes named weight prediction, is vital in exploring the diversity of connectivity pattern emerged in networks. Due to the fundamental significance, it has drawn much attention in the field of network analysis and mining. Some related works appeared in recent years have significantly advanced our understanding of how to predict the strong and weak ties in the social networks. However, most of the proposed approaches are scenario-aware methods heavily depending on some special contexts and even exclusively used in social networks. As a result, they are less applicable to various kinds of networks.

In contrast to the prior studies, here we propose a new computational framework called Neighborhood Estimating Weight (NEW) which is purely driven by the basic structure information of the network and has the flexibility for adapting to diverse types of networks. In NEW, we design a novel index, i.e., connection inclination, to generate the representative features of the network, which is capable of capturing the actual distribution of the tie strength. In order to obtain the optimized prediction results, we also propose a parameterized regression model which approximately has a linear time complexity and thus is readily extended to the implementation in large-scale networks. The experimental results on six real-world networks demonstrate that our proposed predictive model outperforms the state of the art methods, which is powerful for predicting the missing tie strengths when only a part of the network's tie strength information is available.
\end{abstract}


\end{frontmatter}


\section{Introduction}
Traditionally, tie strength is a purely sociological concept which is used to describe how close the relationships are between people in a social network \cite{burke2014growing}. In this study, we expand its connotation to address the connection strength in any kinds of networks. If the tie strength could be quantified by a value, a larger value would represent a tighter connection called a strong tie and a smaller one corresponds to a looser relationship called a weak tie. It is worth mentioning that the true meaning of the tie strength would be different on a case-by-case basis and highly depend on the specified scenario. For example, the friendships in social networks have some apparent difference between close friends and mere acquaintances, implying the degree of intimacy between two friends can be regarded as the "tie strength" of friendship. In an email communication network, the frequency of email exchanges would be able to account for the "tie strength" between the sender and the recipient, i.e., the more frequent email contacts represent the stronger tie strength.

In the field of network mining and analysis, tie strengths actually play an important role in reflecting the pattern of connections between the memberships in a network when we explore the mechanism of link formation, the evolution of the community and the growth of the network \cite{kossinets2006empirical}. Yet, in some cases, part of the tie strengths would be invisible due to the information is missing or undetectable. So, it is of theoretical interest to restore or infer the unobservable tie strengths based on the known information in a network. How to accurately predict the tie strength in various networks is the main challenge in this study. Because the tie strength is commonly marked as weight over an edge when we formulate a network into a weighted graph, the tie strength prediction can be also regarded as a problem of weight prediction.

As aforementioned, the tie strengths would represent various meanings from distinct contexts. As a result, most proposed tie strength prediction models highly rely on the use of semantic features such as the profile information of the on-line users which are strongly correlated with the scenario. This means that the model would be effective in a context but fail to be applicable in a different context. Due to lack of generality for most existing tie strength prediction models, we are motivated to study a generic model which can be applied to fulfill the task of tie strength prediction in many kinds of networks. As most networks can be represented by the weighted graphs, we try to merely use the topological information of the graph to generate the features and design the predictive model while discarding the traditional idea of taking in account any possibly available semantic information existing in the networks. By doing so, the generality of the proposed model will be guaranteed.

The innovations and contributions of our study to tie strength prediction are three-fold.\\
(i) We proposed a generic computational framework by simply using Neighborhood information to Estimate the Weights (NEW) of links in the network. Extensive experiments conducted on real world networks verify that the model has a better performance in prediction accuracy against the state of the art methods.\\
(ii)The learning model approximately has a linear time complexity for its computations mainly rely on the magnitude of the links in the networks, which means that the model can be readily scaled up to handle networks with very large size in a nearly linear growth of time consumption .\\
(iii) To figure out the role of the free parameter introduced in the learning model, we performed some comprehensive analyses and found that the parameter tuning actually can statistically allow the predicated weights having a closer center as well as a more similar distribution to the actual weights.

The rest of the paper has a layout organized as follows.
In section 2, we will review the related studies in this field. Then, we will clearly define the problem of tie strength prediction and introduce the baseline methods in section 3. In the next section, we will investigate how to exploit the network features by using neighborhood information of the nodes. Further, we proposed a learning model to predict the unobserved tie strength. In section 5, we present the experimental results including the accuracy comparisons with some baseline methods. In section 6, some analyses on the proposed model are conducted. In the last section, we conclude our work and give some outlooks of the future works.

\section{Related work}
In this paper, our work is relevant to the tie strength determination and the study of the predictive model. Therefore, we mainly review some related literature from the two aspects.
\subsection{The background knowledge of the tie strength}
In Granovetter's pioneering work \cite{granovetter1977strength}, he had initially addressed the conception of tie strength in social relationships among people and highlighted the role of weak tie in the spread of information. In particular, Granovetter characterized the tie strength from four dimensions including \textit{amount of time}, \textit{intimacy}, \textit{intensity} and \textit{reciprocal services}. Thanks to the rich on-line features provided by the Facebook, Gilbert et al. \cite{gilbert2009predicting} widened the scope of featuring the tie strength to seven dimensions with 74 variables. From the previous studies, to the best of our knowledge, tie strength would be a sociological notion mainly used in the analysis of human relationships. In this work, we borrow this concept as a unified statement to address any connection strength between entities in diverse networks like strength of interaction between proteins in a biological network.

On the other hand, how to obtain the ground-truth data of the tie strength in a given network is still an open problem. According to the literature \cite{marsden1984measuring,rotabi2017detecting,moore2013beware}, there are roughly three ways to fulfill this task. The first one is to collect feedbacks from the users who agree to participate in a survey regarding the degree of relationships with their friends which is initiated by the researchers. This is the most common method to get the social tie strength between people. The second one is to make use of a trusted network to determine strong ties. For example, the phone book network is considered as a trusted network and the tie strength between people who have phone contact with each other is recognized as the strong tie. Thirdly, in some other works, researchers define the tie strength in terms of its actual meaning in the network. One of the typical examples is the rating scores between users in the Bitcoin network, which represent the degree of trusting each other between the buyer and the seller in a transaction of the virtual coins. As the networks studied in this work are not limited to the social networks, we adopt the third way to define the tie strength in the networks for its adaptability to diverse networks. 
\subsection{Overview of the studies of tie strength prediction}
As a branch of the link prediction, the study of tie strength prediction has received much attention from researchers in recent years and some interesting models regarding this topic are proposed \cite{gilbert2012predicting,marsden2012reflections}. Here, we need to point out that most of them prefer to treat it as a binary prediction problem, namely classification of the strong and weak ties. For example, Xiang et al. proposed a model to infer relationship strength based on profile similarity and interaction activity, with the goal of automatically distinguishing strong relationships
from weak ones \cite{xiang2010modeling}. Kahanda \& Neville \cite{kahanda2009using} applied a supervised learning approach to the problem by constructing a predictor that determines whether links in an online campus network are the strong or weak ties. They report that the network's transactional features (such as communications or file transfers) are the most influential features for this task. Rotabi et al. tried to use motifs (subgraphs frequently appeared in a network) on multiple networks to detect strong ties \cite{rotabi2017detecting}. The closed triadic structure previously was deemed quite useful for link prediction in the social networks\cite{leskovec2010signed,leskovec2010predicting}. But, in this work, they found some motifs being larger than the triadic structure are even better in the context of strong tie prediction. Jones et al. used online interaction data (interactions on Facebook like comments, messages, wall post, etc.) to successfully identify real-world strong ties between some surveyed Facebook users via using classic classifiers including logistic regression, support vector machine and random forest \cite{jones2013inferring}.

Note that, almost all the mentioned works prefer to utilize the semantic features rather than the structural features or combine the two sides together in their models. Although the predictive model proposed by the Rotabi et al. is based upon the network structure, it focused on the tie strength prediction in multiple networks and additionally used a reference network with weak ties as the priori knowledge. For a given weighted network, we argue that the graph structure itself has already provided us the sufficient and useful information which can be utilized for modeling the tie strength prediction. Moreover, an advantage of merely using the structural information of the ego network can ensure the predictive model generic since it is free of the dependence on any particular contexts.
 
In addition, unlike the problem setting in previous works, our study does not simply treat the tie strength prediction as a binary classification problem. Instead, we formulate it as an issue of multi-value prediction which would be even more challenging. Similar to the rating prediction problems defined in the study of recommender system \cite{koren2009collaborative,liu2018learning}, we define our prediction task as an explicit tie strength prediction while the task of binary classification for strong and weak ties belongs to the implicit tie strength prediction.


At the same time, there is an another active branch which is working on the tie strength prediction in weighted signed networks (WSNs) \cite{leskovec2010predicting,tang2016survey}. The weighted signed networks refer to networks having edges labeled with positive and negative weights. Kumar et al. proposed two novel measures to quantify the user's behaviors including so-called $goodness$ and $fairness$ and used them to predict the weights in the WSNs \cite{kumar2016edge}. Wang et al. proposed a novel and flexible end-to-end Signed Heterogeneous Information Network Embedding (SHINE) framework to extract users’ latent representations from heterogeneous networks and predict the sign of unobserved sentiment links \cite{wang2018shine}. Despite that signed edges are not necessary for our problem setting, their research ideas are instructive and inspiring to our study.

To the best of our knowledge, we finally only find three works which have the similar motivation as our paper to study the explicit prediction of the tie strength by merely using the network's topological characteristics. In Lv et al.'s study \cite{lu2010link}, they uncovered that weak ties play a significant role in the problem of link prediction in terms of the motif analysis made in some empirical networks and defined three weighted structural similarity indices which can be used for tie strength prediction. The second is the study conducted by Zhao et al. \cite{zhao2015prediction} who also proposed three weighted local measures based on the reliable-route to perform the tie strength prediction. The last one is found that Fu et al. \cite{fu2018link} proposed the line graph indices by converting the edges to the nodes and utilizing the node centrality measures in line graph to define the importance of links in original graph. The approaches proposed by the three highly related works are adopted for comparisons with the new model to verify its superiority. The comparison results will be presented in detail in section 5.

\section{Preliminaries}
In this section, to clearly address the problem of tie strength prediction studied in this paper, an example is introduced to explain the steps of the whole process. Meanwhile, aiming to this issue, some typical topology-based methods for tie strength prediction without considering any concrete scenarios are briefly introduced here as well.

\subsection{Problem description}
\begin{figure}[htbp]
\centerline{\includegraphics[width=4in]{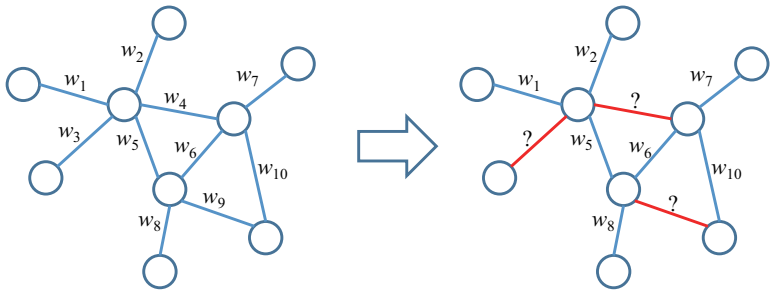}}
\caption{(a) A weighted toy network. (b) The network with weights removed from three edges marked by red color.}
\label{fig:1}
\end{figure}

A weighted network can be represented by a 3-tuple $G=(V, E, W)$ where $V$ denotes the set of the nodes, $E$ denotes the set of the edges or links and $W$ denotes the set of the weights ( each $w\in \mathbb{R}^+$) over edges which reflect the tie strengths of the links. Here, we give a toy example to illustrate the problem of tie strength prediction. The Fig. \ref{fig:1}(a) shows a complete weighted network $G$. If we randomly remove a fraction of weights $W^\ast$ (e.g., 10\% of weights in the network) but keep the edges $E^\ast$ marked by red color in the network as shown in Fig. \ref{fig:1}(b), the task of the tie strength prediction for the network $G^\ast=(V,E,W-W^\ast)$ is to restore the removed weights $W^\ast$ over the edges $E^\ast$ as accurately as possible. In practice, for a network with missing weights, one can use the current network as an observable network or a training network and develop a predictive model to predict the weights absent from the links.

\subsection{Similarity based methods}
As mentioned in the former section, only a few of the tie strength prediction approaches are independent to the scenario and mainly based on the calculations of the network's local structure similarity. In a weighted graph, the two ends of an edge is called a node pair. For a given connected node pair $(x,y)$ with missing tie strength, the calculated similarity between the two nodes can be directly used as an estimation of the tie strength. Here, some commonly used similarity measures for tie strength prediction are introduced and will be treated as baseline methods for comparison in the section of experimental results.

(1) Weighted Common Neighbor (WCN) measure:
\begin{equation}
S_{xy}=\sum_{z \in {N(x) \cap N(y)}}{(w_{xz}+w_{zy})},
\label{eq1}
\end{equation}
where $N(x)$ denotes the neighbors of node $x$.

(2) Weighted Adamic Adar (WAA) measure:
\begin{equation}
S_{xy}=\sum_{z \in {N(x) \cap N(y)}}{\frac{(w_{xz}+w_{zy})}{\log{(1+s(z))}}}.
\label{eq2}
\end{equation}

(3) Weighted Resource Allocation (WRA) measure:
\begin{equation}
S_{xy}=\sum_{z \in {N(x) \cap N(y)}}{\frac{(w_{xz}+w_{zy})}{s(z)}}.
\label{eq3}
\end{equation}
The $s(z)$ shown in Eqs. (\ref{eq3}) and (\ref{eq4}) is the sum over weights of links attaching to node $z$ which has a form as
\begin{equation}
s(z)=\sum_{i \in {N(x)}}{w_{iz}}
\label{eq4}
\end{equation}
(4) The reliable-route Weighted Common Neighbor (rWCN) measure:
\begin{equation}
S_{xy}=\sum_{z \in {N(x) \cap N(y)}}{(w_{xz}.w_{zy})},
\label{eq5}
\end{equation}
where $N(x)$ denotes the neighbors of node $x$.

(5) The reliable-route Weighted Adamic Adar (rWAA) measure:
\begin{equation}
S_{xy}=\sum_{z \in {N(x) \cap N(y)}}{\frac{(w_{xz}.w_{zy})}{\log{(1+s(z))}}}.
\label{eq6}
\end{equation}

(6) The reliable-route Weighted Resource Allocation (rWRA) measure:
\begin{equation}
S_{xy}=\sum_{z \in {N(x) \cap N(y)}}{\frac{(w_{xz}.w_{zy})}{s(z)}}.
\label{eq7}
\end{equation}
It is worth mentioning that, as a step of data pretreatment, one needs to map the weights into the range of (0,1) by using the following formula before calculating the above six similarity indices. 
\begin{equation}
w'=e^{-\frac{1}{w}}.
\label{eq8}
\end{equation}

\section{The computational framework of NEW}

\subsection{Implicit structure feature mining}
\begin{figure}[htbp]
\centerline{\includegraphics[width=2in]{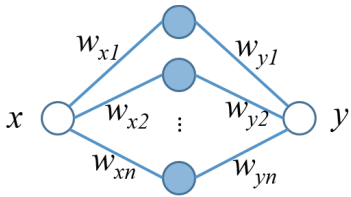}}
\caption{Illustration of the common neighbors or 2-order paths between the nodes $x$ and $y$.}
\label{fig:2}
\end{figure}
Inspired by the theory of homophily from sociology \cite{krivitsky2009representing} (i.e., people are more likely to form strong ties when they share more attributes such as tweets or followers in an on-line social network), we focus on the node pair's  common neighbors to exploit the useful local structure features which would potentially affect the tie strength. In light of this idea, as shown in Fig. \ref{fig:2}, we consider that the existing common neighbors actually have shown us how many connections with 2-order paths exist between the nodes $x$ and $y$. It is reasonable for us to pay attention to these paths because they are supposed to provide us some indirect evidences or clues about the tie strength between the two nodes. Here, we introduce a local structural feature based upon the common neighbors to reflect the tendencies of two nodes connecting to each other. The feature has a computable expression for each end of the node pair as
\begin{equation}
r_x=\sum_{i \in {N(x) \cap N(y)}}\frac{w_{xi}}{w_{xi}+w_{yi}},
r_y=\sum_{i \in {N(x) \cap N(y)}}\frac{w_{yi}}{w_{xi}+w_{yi}},
\label{eq9}
\end{equation} 
We call the feature as connection inclination which quantifies, for a node pair $(x,y)$, how strong one node tends to connect to the other. We infer that there are three cases for the node pair including both $x$ and $y$ having strong connection inclinations, one node having strong connection inclination yet the other node having weak connection inclination, and both $x$ an $y$ having weak connection inclinations. So, from a perspective of network structure, the tie strength would be simultaneously impacted by the connection inclinations from the two ends of the node pair. We will analyze the significance of the connection inclination feature in Section of model analysis.

\subsection{Parameterized regression model}
In this section, we consider to propose a function to fit the tie strength by using the exploited structural features from an observable weighted network. First of all, we formulate the task of the tie strength prediction into a regression problem which has a mathematical form as
\begin{equation}
L(W,X,\Theta)=\|W-f(X,\Theta)\|_2^2,
\label{eq10}
\end{equation}
where $W=(w_1,w_2,...,w_M)$ is a vector containing the weights of $M$ observed ties in the network, $f(X,\Theta)$ is a fitting function to ensure the best fit to the weight vector. In the fitting function, a $M\times2$ attribute matrix $X$ corresponds to the node pairs of the ties in the network, in which the entries of the $i$th row of the matrix, i.e., $x_{i1}$ and $x_{i2}$, contain the values of the $i$th node pair's structural attributes, namely the calculated connection inclinations for the two ends of the node pair. $\Theta$ is a row vector with unknown parameters, and $\|\cdot\|_2$ denotes the $L_2$ norm. The objective function of the regression problem is to minimize the loss function as
\begin{equation}
\mathop{\arg\min}_{\Theta}L(W,X,\Theta)=\mathop{\arg\min}_{\Theta} \|W-f(X,\Theta)\|_2^2.
\label{eq11}
\end{equation} 
To ensure the constant parameter can be included in the model training, we further add a column of ones to define the generalized form of the feature matrix $X$ as
\begin{equation}
X^*={
	\left[ \begin{array}{ccc}
	x_{11} & x_{12} & 1\\
	x_{21} & x_{22} & 1\\
	... & ... & ...\\
	x_{M1} & x_{M1} & 1
	\end{array} 
	\right ]}.
\label{eq12}
\end{equation}
To fit the weights of ties in the training network, we define a fitting function as
\begin{equation}
f(X^*,\Theta)=(X^*)^{\circ k}\Theta^T,
\label{eq13}
\end{equation} 
where the operator $\circ k$ refers to the $k$ Hadamard power \cite{styan1973hadamard} and $\Theta^T$ means the transpose of the vector $\Theta$. Substituting Eq. \ref{eq13} into Eq. \ref{eq11}, we can get an expression as
\begin{equation}
\begin{aligned}
&\mathop{\arg\min}_{\Theta}L(W,X^*,\Theta)=\\ &\mathop{\arg\min}_{\theta_1,\theta_2,\theta_3} \sum_{i}{\frac{1}{2}(w_{i}-(\theta_{1}x_{i1}^k+\theta_{2}x_{i2}^k+\theta_3))^2}
\label{eq14}
\end{aligned}
\end{equation}
where $\Theta=(\theta_1, \theta_2, \theta_3$) is a vector of three unknown parameters to be determined and $k$ is a free parameter which would have varied values along with different networks. To obtain the optimal parameters, we need to solve the above optimization problem. Meanwhile, to avoid the issue of over-fitting possibly existing in the learning model, we add a regulation term by slightly changing the Eq.(\ref{eq14}) as
\begin{equation}
\begin{aligned}
&\mathop{\arg\min}_{\Theta}L(W,X^*,\Theta,\lambda)=\\ &\mathop{\arg\min}_{\theta_1,\theta_2,\theta_3}
\sum_{i}{\frac{1}{2}(w_{i}-(\theta_{1}x_{i1}^k+\theta_{2}x_{i2}^k+\theta_3))^2}
+\frac{1}{2}\lambda\|\Theta\|_2^2
\label{eq15}
\end{aligned}
\end{equation}
where $\lambda$ is a constant commonly set as $1$. To solve Eq.(\ref{eq15}), we use the technique of stochastic gradient descent (SGD) \cite{vzilinskas2006practical} to calculate the parameters of $\Theta$. The partial derivatives of the Eq.(\ref{eq15}) on the $\Theta$ are
\begin{equation}
\begin{aligned}
\frac{\partial L}{\partial \theta_1}&=-\sum_{i}{(w_{i}-(\theta_1x_{i1}^k+\theta_2x_{i2}^k+\theta_3))x_{i1}^k}+\lambda\theta_1
\label{eq16}
\end{aligned}
\end{equation}
\begin{equation}
\begin{aligned}
\frac{\partial L}{\partial \theta_2}&=-\sum_{i}{(w_{i}-(\theta_1x_{i1}^k+\theta_2x_{i2}^k+\theta_3))x_{i2}^k}+\lambda\theta_2
\label{eq17}
\end{aligned}
\end{equation}
\begin{equation}
\begin{aligned}
\frac{\partial L}{\partial \theta_3}&=-\sum_{i}{(w_{i}-(\theta_1x_{i1}^k+\theta_2x_{i2}^k+\theta_3))}+\lambda\theta_3
\label{eq18}
\end{aligned}
\end{equation}
Thereby, we are able to update the parameters by calculating the Eq.(\ref{eq19}) iteratively until they reach the state of convergence.
\begin{equation}
\Theta_{new}=\Theta_{old}-\alpha\nabla_{\Theta}L(W,X^*,\Theta,\lambda)
\label{eq19}
\end{equation} 
where $\alpha$ is a learning rate which commonly has a value in the scope of $(0,1)$. Some detailed implementations of the model are formally described in Algorithms 1 and 2.
\renewcommand{\algorithmicrequire}{\textbf{Input:}}
\renewcommand{\algorithmicensure}{\textbf{Output:}} 
\begin{algorithm}
	\caption{Training of the parameterized regression model }
	\label{alg:A}
	\begin{algorithmic}[1]
		\REQUIRE ~~\\
		$G^*$: observed network;\\
		$k$, $\alpha$, $\varepsilon$: hyper-parameter, learning rate and threshold value for convergence;
		\ENSURE ~~\\
		$\Theta$: parameter vector;\\
		\STATE {//Initializing the attribute values for all node pairs with observed tie weights in the training network}
		\FOR {i=1 to M}
		\IF{the $i$th node-pair $(x,y)$ has common neighbors in $G^*$}
		\STATE {$x_{i1} \gets \sum_{j \in {N(x) \cap N(y)}}\frac{w_{xj}}{w_{xj}+w_{yj}}$}
		\STATE {$x_{i2} \gets \sum_{j \in {N(x) \cap N(y)}}\frac{w_{yj}}{w_{xj}+w_{yj}}$}
		\ENDIF
		\ENDFOR
		\STATE {$X^* \Leftarrow  X$ //Constructing attribute matrix $X^*$}
		\STATE {//Training the parameters via SGD}
		\STATE {$\Theta_{old} \gets (0,0,0), \Theta_{new} \gets (1,1,1)$}
		\WHILE {$\lVert\Theta_{new}-\Theta_{old}\rVert_2>\varepsilon$ }
		\STATE {$\Theta_{old} \gets \Theta_{new}$}
		\STATE {$\nabla L(\Theta) \gets \frac{\partial L(W,X^*,\Theta,\lambda)}{\partial \Theta}$ //Using Eqs. (\ref{eq16})-(\ref{eq18}) to calculate the partial derivatives of the parameters}
		\STATE {$\Theta_{new} \gets \Theta_{old}-\alpha\nabla L(\Theta)$ //Updating parameter vector}
		\ENDWHILE
		\RETURN $\Theta_{new}$
	\end{algorithmic}
\end{algorithm}

\renewcommand{\algorithmicrequire}{\textbf{Input:}}
\renewcommand{\algorithmicensure}{\textbf{Output:}} 
\begin{algorithm}
	\caption{Tie strength prediction}
	\label{alg:A}
	\begin{algorithmic}[1]
		\REQUIRE ~~\\
		$E^*$: links in $G^*$ with weights removed;\\
		$k$: hyper-parameter value;\\
		$\Theta$: the parameter vector obtained by Algorithm 1;\\
		\ENSURE ~~\\
		$\hat{W}$: the predicted weight vector;\\
		\FOR { $i\in E^*$ }
		\STATE{ $\hat{w^i}_{xy}=\theta_1(r^i_x)^k+\theta_2(r^i_y)^k+\theta_3$ }
		\STATE{//Calculating the weight value for each link in the test set}
		\ENDFOR
		\RETURN $\hat{W}$
	\end{algorithmic}
\end{algorithm}

\section{Experimental Results}
\subsection{Data description}
To evaluate the performance of the proposed model, six real-world networks are used for the tests. Some background information of the networks as well as the specified definitions of the tie strength are briefly introduced as follows.
\begin{itemize}
\item $Caenorhabditis\ elegans$ \cite{liu2013correlations}: This is a metabolic network of the roundworm Caenorhabditis elegans. Nodes are metabolites (e.g., proteins), and edges are interactions between them which are undirected. There may be multiple interactions between any two metabolites, representing the tie strength between them.

\item $UC\ social\ network$ \cite{opsahl2009clustering}: This directed network contains sent messages between the users of an online community of students from the University of California, Irvine. A node represents a user. A directed edge represents a sent message. The weights denote multiple messages. Here, we simply treat it as an undirected and weighted graph .

\item $Political\ blogs$ \cite{adamic2005political}: A network with directed hyperlinks between political web blogs on 2004 US Election is recorded in 2005 by Adamic and Glance. We treat it as an undirected graph and the weight for an edge is defined by the multiple hyperlinks between two blogs.

\item $Coauthorships\ in\ network\ science$ \cite{newman2006finding}: A co-authorship network of scientists working on network theory and experiment, as compiled by M. Newman in May 2006. Here, we merely use the largest component of this network and the weights represent the strength of the collaboration  between scientists marked by M. Newman which are calculated via the information of co-authored papers and the number of the authors.

\item $Neural\ network$ \cite{watts1998collective}: A directed, weighted graph represents the neural network of C. Elegans, in which an edge joins two neurons if they are connected by either a synapse or a gap junction. The network is treated as the undirected graph and the weights on edges denote the multiple interactions between the neurons.


\item $German\ Wikipedia$ \cite{margaretha2014building}: This is a dataset of discussion threads on the German Wikipedia. Nodes of the network are users of the German Wikipedia. A directed link from user A to user B denotes that user A wrote a comment in a discussion as a reply to a comment of user B. Multiple comments between the users A and B form the tie strength which can be marked as the weight over the edge between A and B.
\end{itemize}

Some basic statistics of the six networks are summarized in Table \ref{tab1}.
\begin{table*}[htbp]
\caption{Statistics of the networks' basic information. \textless w\textgreater \, denotes the average of the weights in the network. Max(w) and Min(w) represents the maximum and minimum of the weights in the network, respectively.}
\begin{center}
\resizebox{\textwidth}{12mm}{
\begin{tabular}{ccccccc}
\hline
 & \multicolumn{6}{c}{\textbf{Networks}} \\
\cline{2-7} 
\textbf{Features} & \textbf{\textit{UC social}}& \textbf{\textit{C. elegans}}& \textbf{\textit{P. blogs}} & \textbf{\textit{Netscience}} & \textbf{\textit{Neural network}} & \textbf{\textit{Wiki}}  \\
\hline
\# of nodes & 1899 & 453 & 1224 & 575 & 296 & 90153 \\
\# of edges & 13838 & 2025 & 16718 & 1028 & 2137 & 727870 \\
\textless w\textgreater & 4.324 & 2.2529 & 1.1419 & 0.3167 & 4.1132 & 2.8181\\
Max(w) & 184 & 114 & 3 & 2.5 & 72 & 1355 \\
Min(w) & 1 & 1 & 1 & 0.0526 & 1 & 1 \\
\hline
\end{tabular}
}
\label{tab1}
\end{center}
\end{table*}

For setting up the experiment, we adopt the scheme of leave-10\%-out prediction which is widely used for model assessment in the domain of link prediction \cite{cannistraci2013link,lu2011link}. The weights from some randomly selected edges in the original network with a fixed proportion of 10\% are removed. Then, the 10\% links as well as the attached weights constitute a probe or test set while the remainder of the network is treated as the training set. Note that the topology information of the entire network, i.e., $(V,E)$, is always available. Thus, a weighted network is properly divided into two parts which allow us to conduct the model parameter training and the test of the tie strength prediction, respectively. We call such a process as one partition of the network. In case of that a single network partition would result in an unexpected biased experimental result, we adopt ten independently random partitions on the original network to generate ten groups of training and probe sets and take the averaged results over the total of ten experiments as a criterion to assess the performance of the model.

\subsection{Metric for evaluation}
The frequently used evaluation metrics in the related literature \cite{fu2018link,zhao2015prediction} include Pearson Correlation Coefficient (PCC) and Root Mean-Squared Error (RMSE). Due to the reason that the mentioned studies have verified both of them are able to consistently evaluate the accuracy of the explicit tie strength prediction,  we merely choose one of two metrics, i.e., the Root Mean-Squared Error, for experimental results evaluation in this work, which has a mathematical form as
\begin{equation}
RMSE=\sqrt{\frac{1}{n}\sum_{i=1}^N{(w_i-\hat{w_i})^2}},
\label{eq20}
\end{equation}
where $W=(w_1,w_2,...,w_N)$ is the actual weight vector for $N$ existing links in the test set and $\hat{W}=(\hat{w_1},\hat{w_2},...,\hat{w_N})$ is the predicted weight vector for the corresponding links. This metric can measure how close the values of the predicted weights and the actual ones are. Therefore, the smaller the value of RMSE is, the better the accuracy of the tie strength prediction.
 
\subsection{Comparisons of the prediction results}
\begin{figure}[htbp]
	\centerline{\includegraphics[width=3.5in]{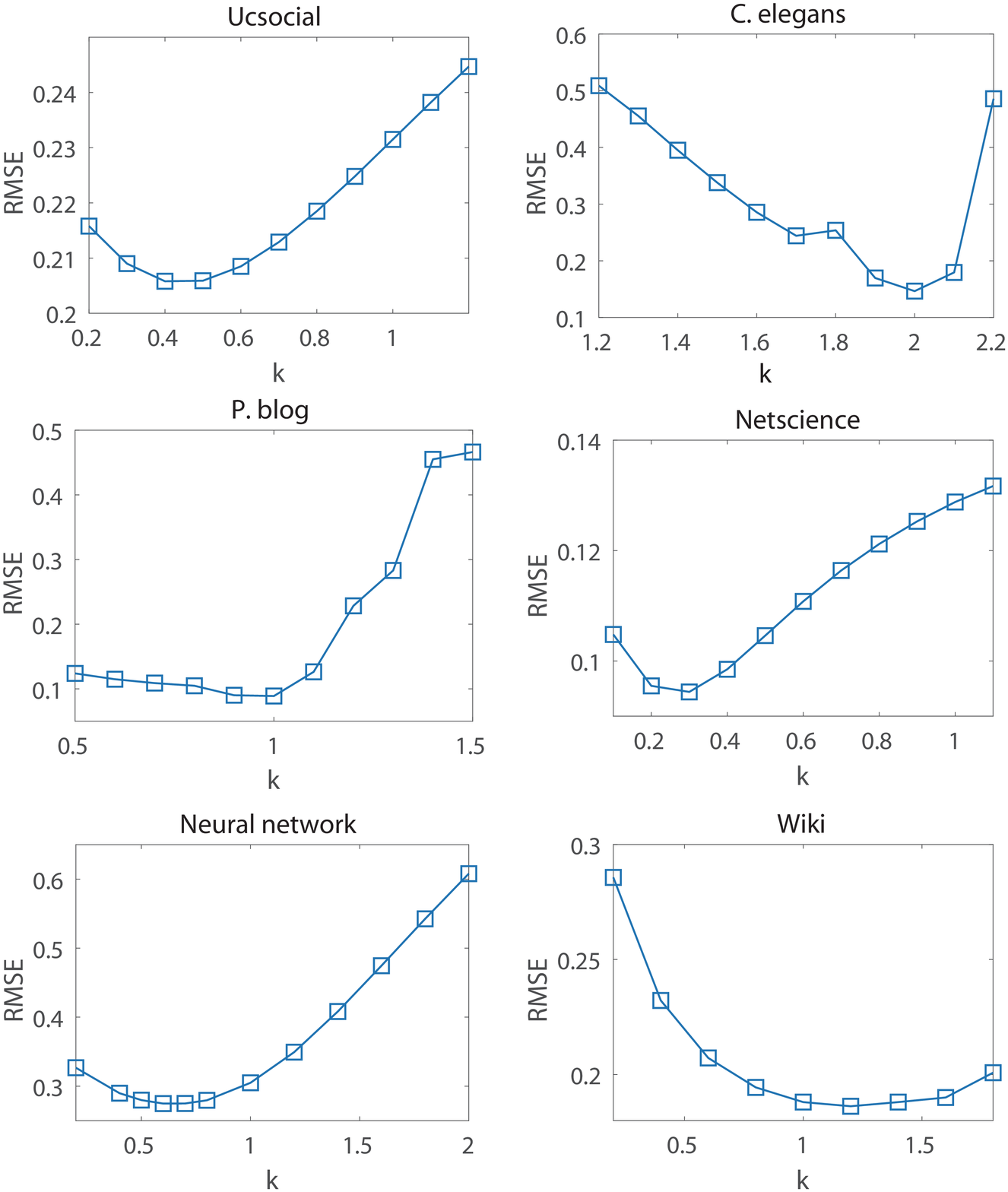}}
	\caption{The average RMSE curves under the tuning of parameter $k$ for the six networks.}
	\label{fig:3}
\end{figure}

Since the $k$ introduced in Eq. (\ref{eq15}) is a hyper-parameter, a process of parameter tuning is required to obtain the optimized model. For a given network, after performing tests on ten pretreated network partitions with the same setting of $k$ value, we average the RMSE values obtained from the ten experiments and can derive the mean value of the RMSEs. Repeating this process by changing the value of $k$, we will derive a series of averaged RMSE values with a variety of $k$ value settings. Putting the obtained RMSE mean values together, we will find out an optimal $k$ value. Apparently, the optimal $k$ value will result in the smallest RMSE mean value. It turns out that the optimal value of the $k$ is commonly in the scope of (0,2] in terms of the experimental results on the six real-world networks and the procedure of the model training will become harder to converge when the $k$ is larger than 2. The corresponding curves of RMSE against $k$ for the six networks are illustrated in Fig. \ref{fig:3}. The optimal $k$ values are 0.4, 2, 1, 0.3, 0.6 and 1.2 for networks of $UC\ social$, $C.elegans$, $P.blog$, $Netscience$, $Neural\ network$ and $Wiki$, respectively.

Here, the machine used for the test is a desktop with 16 Gigabyte memory and 3.0 GHz Intel core-i3 processor. The results reported in Table \ref{tab2} are the RMSE mean values as well as their standard deviations over ten experimental results obtained by the NEW model and nine baseline methods on six real-world networks. Among the nine compared approaches, six indices have been introduced in Section 3.2 while the three other ones are line-graph based approaches. The main idea of line-graph model is to calculate node centrality with 8 Centrality-based measures such as Closeness Centrality, Betweenness Centrality, and Eigenvector Centrality, etc., in the line graph and convert them into 8 features of the edges in the original graph. Note that a node in the line graph is actually corresponding to an edge in the original graph.  Thus, one can apply the edge features into some learning models like Random forest, SVM and GBDT and infer the link weight. The detailed description of line-graph model can be found in literature \cite{fu2018link}. The outcomes shown here for NEW model are calculated with the optimized $k$ settings. Overall, it is verified that our model performs better than the compared methods on the six tested networks. More specifically, compared to the second best method, the proposed model has achieved further declines of the RMSE by 5.19\%, 4.21\%, 16.49\%, 48.32\% 5.09\% and 97.93\% on the networks of $UC\ social$, $C.elegans$, $P.blogs$, $Netscience$, $Neural\ network$ and $Wiki$, respectively. As for network $Wiki$, we failed to predict the link weight via line-graph based approaches due to the Centrality-based measures are too expensive to calculate on a large-sized network with our testing machine. Therefore, aside from having the relatively weaker predictive performance compared to our NEW model, the line-graph based methods are less attractive for owning another significant drawback, i.e., lack of computational scalability.

\begin{table*}[htbp]
\caption{Experimental results on prediction accuracy for the six networks. The best result for each network is marked in boldface.}
\begin{center}
\resizebox{\textwidth}{12mm}{
\begin{tabular}{ccccccc}
\hline
 & \multicolumn{6}{c}{\textbf{Accuracy (RMSE)}} \\
\cline{2-7} 
\textbf{Methods} & \textbf{\textit{UC social}}& \textbf{\textit{C. elegans}}& \textbf{\textit{P. blogs}} & \textbf{\textit{Netscience}} & \textbf{\textit{Neural network}} & \textbf{\textit{Wiki}} \\
\hline
WCN & 8.41$\pm$0.91 & 5.05$\pm$0.55 & 16.65$\pm$0.18 & 0.39$\pm$0.04 & 6.19$\pm$0.85 & 46.17$\pm$0.41 \\
WAA & 8.03$\pm$1.12 & 1.71$\pm$0.48 & 4.61$\pm$0.11 & 1.16$\pm$0.06 & 5.64$\pm$0.75 & 10.61$\pm$0.54 \\
WRA & 8.42$\pm$0.98 & 2.33$\pm$0.76 & 0.94$\pm$0.03 & 1.11$\pm$0.05 & 6.41$\pm$0.75 & 9.23$\pm$0.72 \\
rWCN & 7.97$\pm$1.01   & 1.79$\pm$0.76 & 3.02$\pm$0.03 & 0.43$\pm$0.05 & 5.78$\pm$0.74 & 15.08$\pm$0.39 \\
rWAA & 8.31$\pm$0.99 & 2.36$\pm$0.84 & 0.89$\pm$0.02 & 0.42$\pm$0.05 & 6.3$\pm$0.75 & 8.78$\pm$0.7 \\
rWRA & 8.48$\pm$0.98 & 2.93$\pm$0.91 & 1.09$\pm$0.01 & 0.42$\pm$0.05 & 6.7$\pm$0.76 & 9.38$\pm$0.71 \\
LG-RF & 0.212$\pm$0.003 & 0.174$\pm$0.002 & 0.097$\pm$0.003 & 0.209$\pm$0.004 & 0.216$\pm$0.006 & - \\
LG-GBDT & 0.358$\pm$0.003 & 0.265$\pm$0.003 & 0.221$\pm$0.004 & 0.176$\pm$0.003 & 0.362$\pm$0.005 & - \\
LG-SVM  & 0.213$\pm$0.002 & 0.143$\pm$0.002 & 0.163$\pm$0.002 & 0.201$\pm$0.003 & 0.217$\pm$0.007 & - \\
NEW & \textbf{0.206$\pm$0.002} & \textbf{0.137$\pm$0.003} & \textbf{0.081$\pm$0.001} & \textbf{0.091$\pm$0.002} & \textbf{0.205$\pm$0.006} & \textbf{0.182$\pm$0.002}\\
\hline

\end{tabular}}
\label{tab2}
\end{center}
\end{table*}

\subsection{Performance analysis}
\begin{figure*}[htbp]
	\centerline{\includegraphics[width=5in]{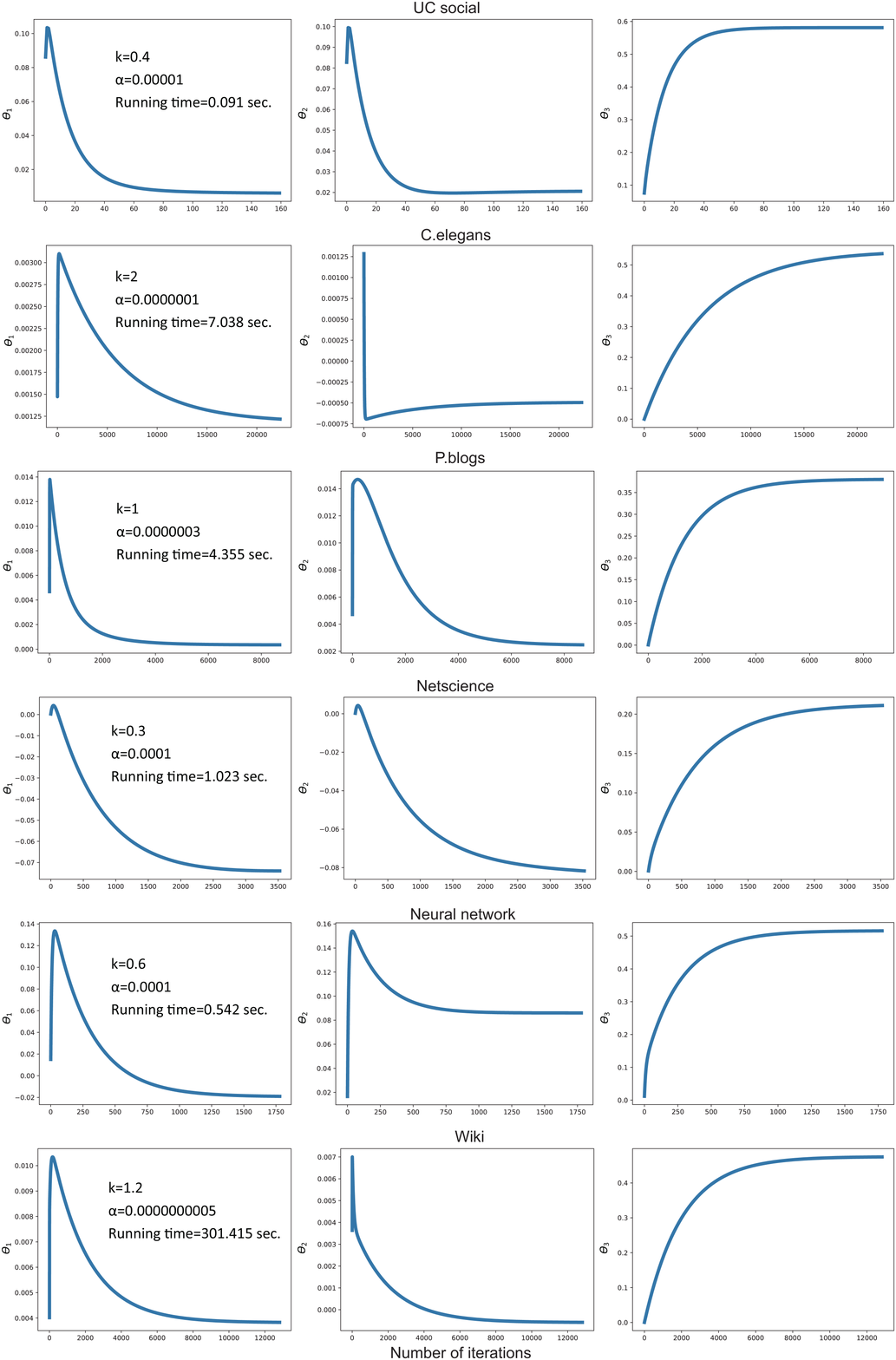}}
	\caption{The convergence curves of parameters $\theta_1$, $\theta_2$ and  $\theta_3$ under model training on one partition for each network. The specified training parameters including hyper-parameter $k$ and learning rate $\alpha$ as well as time consumption for model training  are presented in the plots. }
	\label{fig:4}
\end{figure*}

Based on the derived optimal $k$ values via parameter tunning for the six networks, Fig. \ref{fig:4} illustrates the convergence curves of parameters $\theta_1$, $\theta_2$ and  $\theta_3$ under model training on one partition for each network. The results of running time suggest that the training of model parameters is quite fast on real-world networks as, even on the large-sized network like $Wiki$, it only requires a couple of minutes to finish the training procedure.

\begin{figure}[htbp]
	\centerline{\includegraphics[width=3in]{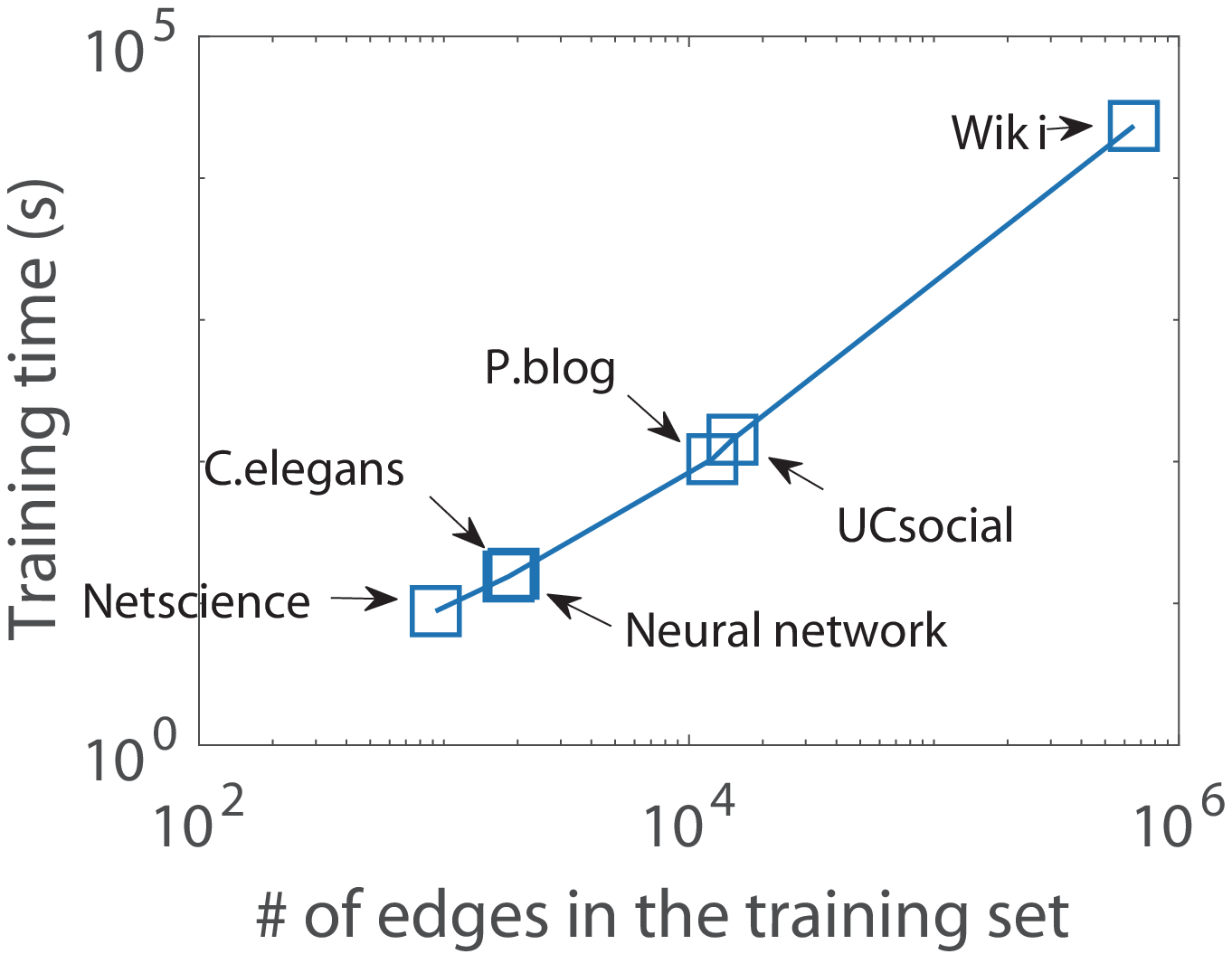}}
	\caption{Scalability of the NEW framework with the sizes of edges in the training sets for the six networks.}
	\label{fig:5}
\end{figure}

According to Algorithm 1, the calculations for the proposed model are mainly related to the number of edges in the training network and the number of common neighbors between node pairs. For initializing the attribute matrix, it requires $O(\hat{N_{cn}}|E-E^*|)$. Due to the fact of sparsity for the real-world network's structure, the averaged common neighbors $\hat{N_{cn}}$ for node pairs are far fewer than the number of the existing edges which can be regarded as a constant. For parameter training, one iteration needs $O(|E-E^*|)$. Yet, the number of iterations required for the parameter convergence depends on the values of $\varepsilon$ and learning rate $\alpha$. Practically, proper settings of the two parameters will lead to a fast convergence of the model which means that the number of the iterations is far fewer than the amount of edges $|E|$. Therefore, we can also treat it as a constant $c$. Thus, the parameter training requires $O(c|E-E^*|)$. As a result, the approximate time complexity of the Algorithm 1 is $O(\hat{N_{cn}}|E-E^*|)+ O(c|E-E^*|)\approx O(|E-E^*|)$. As for Algorithm 2, the time complexity apparently is $O(|E^*|)$. Therefore, Algorithms 1 and 2 have the combined complexity of $O(|E|)$ which is merely correlated with the size of the edges in the network. It proves that our model has very good performance which can be easily scaled up to handle very large sized networks with a linear time complexity. The accumulated training time on the ten training sets for the six networks shown in Fig. \ref{fig:5} also verified that the training time nearly grows linearly along with the size of the training sets.

\section{Model analysis}

\subsection{The significance of the structure features}
\begin{figure}[htbp]
	\centerline{\includegraphics[width=3.5in]{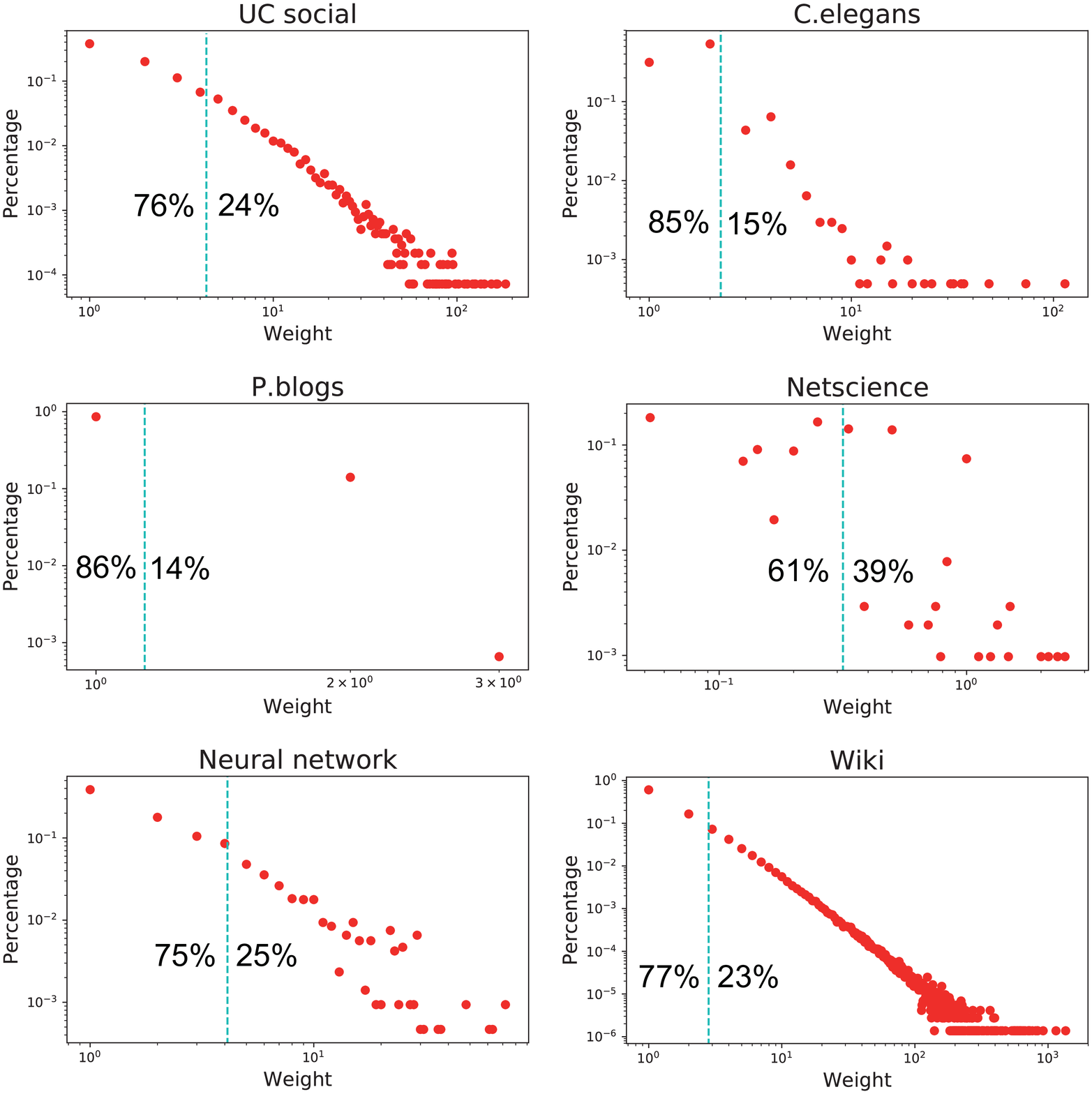}}
	\caption{Weight distributions for the six networks. The dashed line corresponds to the average value of the weights in each plot. The value shown on the left side of the line is the percentage of the small weights and that shown on the right side of the line is the percentage of the large weights. }
	\label{fig:6}
\end{figure}

\begin{figure*}[htbp]
	\centerline{\includegraphics[width=5.5in]{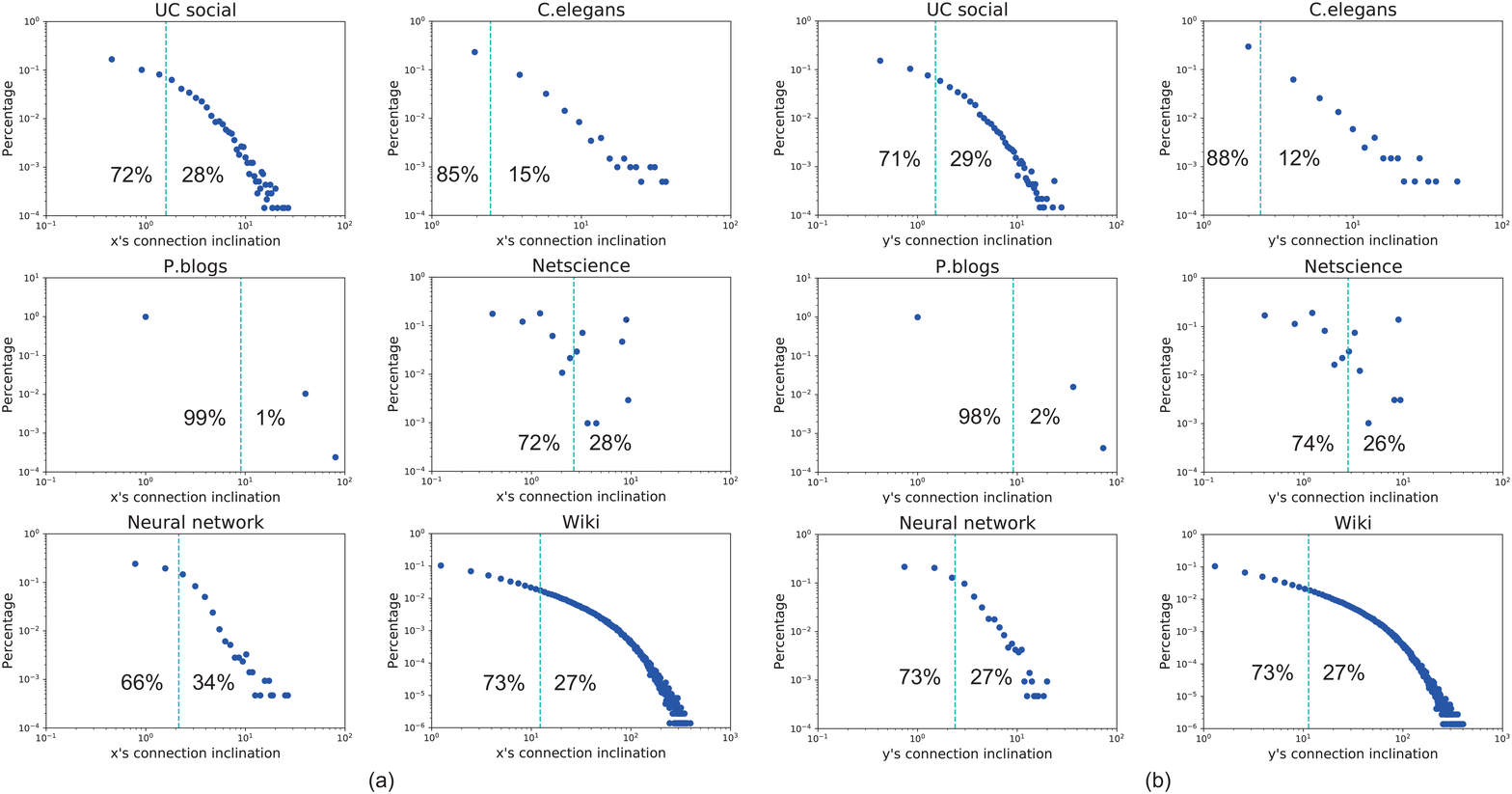}}
	\caption{(a) $x$'s connection inclination distributions for the six networks where $x$ denotes one end of each link. (b) $y$'s connection inclination distributions for the six networks where $y$ denotes the other end of each link. The dashed line corresponds to the average value of the connection inclinations in each plot. The value shown on the left side of the line is the percentage of the small weights and that shown on the right side of the line is the percentage of the large weights.}
	\label{fig:7}
\end{figure*}

Before investigating why the proposed structure features are crucial to our model, we firstly analyze the weight distributions of the six networks as plotted in Fig. \ref{fig:6}. If we take those weights smaller than the average value over all weights in a network as $small\ weight$ and other weights larger than the average value as $large\ weight$, The distributions for different networks show a commonality that weights with relatively small values belong to the majority whereas the large ones are the minority. In particular, weights in the networks of $UC\ social$ and $Wiki$ even have the typical power law distribution which means that nearly 80\% of the weights are $small\ weight$ and 20\% of them belong to the $large\ weight$. That is to say, the real world networks have heavily imbalanced weight distributions.
For each link denoted by $(x,y)$, we can use Eq. (\ref{eq9}) to calculate the dual connection inclinations from both ends of the $(x,y)$. So, we will be able to obtain two distributions of the connection inclinations from all links shown in Fig. \ref{fig:7}. Interestingly, both of the distributions for the six networks are quite similar to the actual weight distributions. Moreover, with regard to the composition of the distributions, even the fractions of the $small\ weight$ and $large\ weight$ for the six networks are very close to their counter parts, i.e., the fractions of the small and large values of connection inclination. It suggests that the proposed two structure features have captured the characteristics of the weight distribution and can be used as good indicators to fit the weights accurately.

\subsection{The role of the parameter $k$ }
\begin{figure*}[htbp]
	\centerline{\includegraphics[width=5.5in]{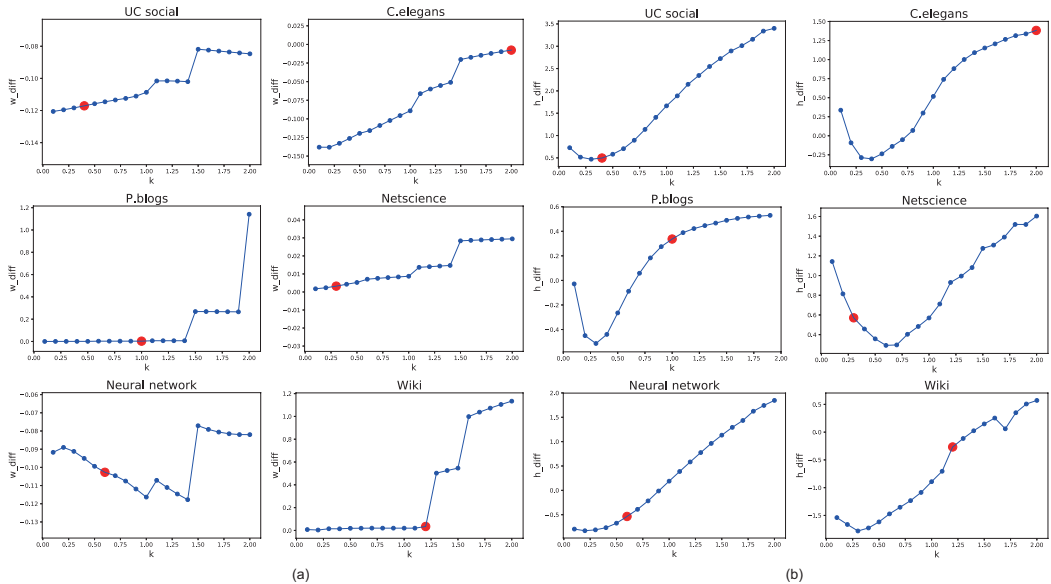}}
	\caption{(a) Perturbation analysis of the $w\_diff$ scores for the six networks. (b) Perturbation analysis of the $h\_diff$ scores for the six networks. The red dot for each plot corresponds to the score of the metric when the parameter $k$ reaches the optimal value for the network. }
	\label{fig:8}
\end{figure*}

In the model design, we initially consider to use a linear regression model. But, in view of that the weights in real world networks commonly having non-uniform distributions, a linear regression model may be useful but not a desired tool to well fit the weights. So, we specially design a non-linear regression model and the objective that we introduce the parameter $k$ into the model is to nonlinearly regulate the process of weight fitting. Via a process of the parameter $k$ tuning, we can obtain the optimized prediction results when an optimal value of the $k$ is determined. Nevertheless, the role of the parameter $k$ is still unclear. Here, we try to analyze this issue from two basic observations. As the amount of the small weights are dominated in the tested networks, most of the predicted weights are supposed to be small as well. This means that the center of the weights and that of the predicted weights would be quite close to each other if we take the mean values averaged over the weights and the predicted ones as the centers of the two distributions, respectively. On the other hand, the weight distributions for all tested networks are heavily imbalanced. As a result, the predicted weights should have the similar or matched distributions if the model has the potential to accurately fit the weights. Here, we introduce two measures to statistically quantify the difference between the real weights and the predicted weights from the two aspects mentioned above. The first measure is to quantify the distance between two distributions' centers.
\begin{equation}
w\_diff=\frac{1}{D}\sum\limits_{j=1}^D(\frac{1}{N}\sum\limits_{i=1}^N w_i-\frac{1}{N}\sum\limits_{i=1}^N \hat{w_i})
\label{eq21}
\end{equation}
where $w_i$ and $\hat{w_i}$ are the weight and the predicted one of the $i$th link among $N$ edges in the test set of the network, respectively. As there are $D$ test sets ($D=10$ in this paper) available due to the multiple independent network partitions, we need to average the $D$ results to ensure the $w\_diff$ score unbiased. The second one is to measure the difference of the distributions using the information entropy.
\begin{equation}
h\_diff=\frac{1}{D}\sum\limits_{j=1}^D(\sum\limits_{i=1}^S p_i\log p_i-\sum\limits_{i=1}^S \hat{p_i}\log \hat{p_i})
\label{eq22}
\end{equation}
where $S$ is the total number of the weight categories in which those weights having identical values represent a category and the ratio of the weight quantity of the $i$th category to the quantity of the total weight is denoted by $p_i$. Accordingly, $\hat{p_i}$ is the percentage of the predicted weights falling in the $i$th category. With respect to the weights and predicted weights, in our opinion,  smaller absolute scores of the $w\_diff$ and $h\_diff$ suggest a better result of the prediction. So, for the six networks, we separately plot the perturbation curves of the $w\_diff$ scores and $h\_diff$ scores under the $k$ tuning as shown in Fig. \ref{fig:8} and mark the corresponding point in red in each plot when $k$ reaches the optimal value. From the Fig. \ref{fig:7}, we observe that the optimal $k$ results in three consequences. Firstly, the optimal $k$ will lead to the least $w\_diff$ score like the networks of $C.elegans$, $P.blogs$ and $Wiki$. Secondly, it will promote the score of $h\_diff$ to be very close to the least value and the networks of $UC\ social$ and $Wiki$ are the examples. Thirdly, it will reach a trade-off state between the scores of $w\_diff$ and $h\_diff$, both of which have the relatively small values such as $Netscience$. We also notice that the scores of $w\_diff$ are impacted more significantly under the parameter tuning when putting the two measures together for comparison. Therefore, we conclude that the tuning of the parameter $k$ mainly promotes the center of the predicted weights approaching the center of the actual weights as closely as possible and also partially ensure the distribution of the predicted weights is similar to the actual distribution of the weights.

\section{Conclusion and future works}
In this paper, we are motivated from a viewpoint of generic learning to study the problem of tie strength prediction in various networks, which is significantly different from the traditional research paradigm mainly focusing on the specified social networks. Following this idea, we do not need to pay attention to any explicit domain features provided by a network which will restrict the utility of the predictive model in a wide range of networks. Instead, we focus on the structure features owned by any kinds of networks. According to the empirical analysis, the proposed index called connection inclination has the capacity to capture the tie strength from both ends of the links which merely uses the shared neighborhood information of the node pair. Aiming to fit the weights having imbalanced distributions in the real world networks, we proposed a parameterized non-linear regression algorithm accordingly. Essentially, comparing to the linear regression model, the non-linear model has a better adaptability to the non-uniform weight distribution. So, both the structure-driven features and non-linear learning model ensure that the entire computational framework is more generic than the conventional methods. The extensive experiments demonstrate that our proposed model overall has better performance than the baseline methods in the accuracy of tie strength prediction on six tested networks. Meanwhile, both the theoretical analysis and time consumption tests on different networks verify that our model has an approximate linear time complexity and thus has a good scalability to the large sized networks. To the best of our knowledge, our work is the first attempt in systematically studying the generic learning model for tie strength prediction in networks which would have provided some insights for the future works in this domain. It is worth noting that the introduced free parameter $k$ plays an important role for the model optimization but technically, parameter tuning would be a tedious task. So, estimation of the optimal $k$ is preferred in practice and also can be treated as a future extension of the current model. Meanwhile, if there is no common neighbors between node pairs, our model simply treats that the features of $r_x$ and $r_y$ equal to 0. In the future extension, we need to handle this case to specially define the features when two nodes of an edge do not possess a common neighbor. It would be also interesting to extend our model to predict the tie strength in temporal networks as the weights of the links would be changing over time. Recently, Qu et al. \cite{qu2019temporal} studied the issue of discovering vital nodes in temporal networks, their findings could enlighten us to exploit some new node features to capture the link weights in time-varying networks.

\section*{Acknowledgement}
 The authors would like to thank the anonymous referees for their valuable comments and helpful suggestions.

\label{}



\bibliographystyle{elsarticle-num}
\bibliography{reference}

\end{document}